\documentclass[12pt]{article}
\usepackage{a4wide}
\usepackage{epsfig}

\newcommand{\slk}{/\kern-6pt k}
\newcommand{\sll}{/\kern-4pt l}
\newcommand{\slp}{p\kern-5pt/}
\newcommand{\slq}{q\kern-5.5pt/}

\newcommand{\oone}{\hbox{$1\kern-2.5pt\hbox{\rm l}$}}
\newcommand{\ssigma}{\hbox{$\kern2.5pt\vrule height4pt\kern-2.5pt\sigma$}}

\newcommand{\GeV}{{\rm\,GeV}}

\begin{document}

\thispagestyle{empty} 
\begin{flushright}
MZ-TH/12-16\\
August 2012
\end{flushright}
\vspace{0.5cm}

\begin{center}
{\Large\bf $O(\alpha_s)$ corrections to the decays of polarized\\[.2cm]
  $W^\pm$ and $Z$ bosons into massive quark pairs}\\[1.3cm]
{\large S.~Groote$^{1,2}$, J.G.~K\"orner$^2$ and P.~Tuvike$^1$}\\[1cm]
$^1$ Loodus- ja Tehnoloogiateaduskond, F\"u\"usika Instituut,\\[.2cm]
  Tartu \"Ulikool, T\"ahe 4, 51010 Tartu, Estonia\\[7pt]
$^2$ Institut f\"ur Physik, Johannes Gutenberg-Universit\"at,\\[.2cm]
Staudinger Weg 7, 55099 Mainz, Germany
\end{center}

\vspace{1cm}
\begin{abstract}\noindent
We present $O(\alpha_s)$ results on the decays of polarized $W^\pm$ and $Z$
bosons into massive quark pairs. The NLO QCD corrections to the polarized
decay functions are given up to the second order in the quark mass expansion.
We find a surprisingly strong dependence of the NLO polarized decay functions 
on finite quark mass effects even at the relatively large mass scale of the
$W^\pm$ and $Z$ bosons. As a main application we consider the decay
$t\to b+W^+$ involving the helicity fractions $\rho_{mm}$ of the $W^+$
boson followed by the polarized decay $W^+(\uparrow)\to q_1\bar{q}_2$
for which we determine the $O(\alpha_s)$ polar angle decay distribution. 
We also discuss NLO polarization effects in the production/decay process
$e^+e^-\to Z(\uparrow)\to q\bar{q}$.
\end{abstract}

\newpage

%%%%%%%%%%%%%%%%%%%%%%%%%%%%%%%%%%%%%%%%%%%%%%%%%%%%%%%%%%%%%%%%%%%%%%%%%%%%%
\section{Introduction} 
%%%%%%%%%%%%%%%%%%%%%%%%%%%%%%%%%%%%%%%%%%%%%%%%%%%%%%%%%%%%%%%%%%%%%%%%%%%%%
The polarization of $W^\pm$ and $Z$ bosons produced in electroweak production 
processes is in general highly nontrivial. One therefore has a rich
phenomenology of polarization effects in $(W,Z)$ production and decay which
will be explored in present and future experiments. The polarization of the
$W^\pm$ and $Z$ bosons can be probed by decay correlations involving the decay
products of the polarized $(W,Z)$ bosons. A widely discussed prominent example
of such decay correlations is the decay $t\to b+W^+$ followed by
$W^+\to\ell^+\nu_\ell$ where the decay $W^+\to\ell^+\nu_\ell$ is used to
analyze the helicity fractions of the $W^+$ resulting from the decay
$t\to b+W^+$ (see e.g.\ Refs.~\cite{Aaltonen:2012rz,Peters:2011an}). It would
be interesting to explore the possibility to also make use of the
quark--antiquark decay modes $W^{\pm}\to q_1\bar q_2$ and $Z \to q \bar q$ to
analyze the polarization of the $(W^{\pm},Z)$ bosons, in particular using the
tagging modes  $W^+\to c\bar b,c \bar s$ and $Z \to c \bar c,b \bar b$ 
involving heavy quarks.

This paper is devoted to the calculation of NLO QCD effects in the decay of 
polarized $(W,Z)$ gauge bosons into massless and massive quark--antiquark 
pairs. In order to provide a quick access to the importance of quark mass
effects we have made use of a quark mass expansion of the rather lengthy fully
analytic NLO results listed in~\cite{gkt12}. We thereby demonstrate that, in
polarized gauge boson decays involving charm and  bottom quarks, the NLO
finite mass effects are non-negligible. The reason is that the NLO finite mass
corrections in polarized decays set in at linear order with rather large
coefficients, contrary to the case of unpolarized decay where the finite mass
effects set in only at $O(m_q^2/m_{W,Z}^2)$. Depending on the particular
polarized decay function, the mass corrections can become as large as the
leading term of the NLO mass expansion for decays involving $b$ and $c$ quarks.

NLO and finite mass effects will affect the decay correlations between the 
momenta of the production and decay process. As specific examples of how such
correlations are affected by NLO and finite mass effects we consider the
cascade decay process $t\to b+W^+(\uparrow)(\to c\bar{b})$ and the
production/decay process $e^+e^-\to Z(\uparrow)\to b\bar{b},c\bar{c}$. 
 
%%%%%%%%%%%%%%%%%%%%%%%%%%%%%%%%%%%%%%%%%%%%%%%%%%%%%%%%%%%%%%%%%%%%%%%%%%%%%
\section{Angular decay distribution}
%%%%%%%%%%%%%%%%%%%%%%%%%%%%%%%%%%%%%%%%%%%%%%%%%%%%%%%%%%%%%%%%%%%%%%%%%%%%%
The polar angle decay distribution of a polarized $(W,Z)$ boson decaying 
into a fermion--antifermion pair (quark or lepton pair) is given by
\begin{eqnarray}
\label{Wtheta}
W(\theta)&=&\sum_{m,m'=0,\pm1}\rho_{mm}\,d^1_{mm'}(\theta)
  \,d^1_{mm'}(\theta)\,\, H_{m'm'}\nonumber\\
  &=&\frac38(1+\cos^2\theta)\,\Big((\rho_{++}+\rho_{--})\,(H_{++}+H_{--})
  +2\rho_{00}H_{00}\Big)\nonumber\\
  &&+\frac34\cos\theta\,\Big((\rho_{++}-\rho_{--})\,(H_{++}-H_{--})\Big)
  \nonumber\\&&
  +\frac34\sin^2\theta\,\Big((\rho_{++}+\rho_{--})H_{00}+\rho_{00}
  (H_{++}+H_{--}-H_{00})\Big)\,.
\end{eqnarray}
The diagonal spin density matrix elements of the polarized $(W,Z)$ boson are
denoted by $\rho_{mm}$.\footnote{In the literature the diagonal elements of
the normalized spin density matrix of the $(W,Z)$ boson $\rho_{mm}$ are
frequently referred to as the helicity fractions ${\cal F}_{m}$ of the gauge 
boson.} They are defined in a $(x,y,z)$ coordinate system associated with the
production process while the polarized decay structure functions (for short:
polarized decay functions) $H_{m'm'}$ are defined in a $(x',y,z')$ coordinate
system associated with the decay process. The two coordinate systems are
rotated into each other by a rotation around the $y$ axis by the polar angle
$\theta$. For example, for the sequential decay
$t\to b+W^+(\uparrow)(\to c\bar{b})$ the $z$ axis could be chosen to lie along
the momentum of the $ W^+$ in the top quark rest frame and the $z'$ axis could
be chosen to lie along the quark direction in the $W^+$ rest frame. 

The polarized decay functions $H_{\pm\pm}$ and $H_{00}$ stand for the
probability of the decay of a polarized vector boson $(W,Z)(m')$ into a
fermion--antifermion pair where the vector boson has the spin quantum numbers 
$m'=0,\pm$ in the rotated $(x',y,z')$ coordinate system. The spins of the
fermion--antifermion pair are summed over when calculating the decay
probability.

We take the spin density matrix to be normalized, i.e.\
$\rho_{++}+\rho_{00}+\rho_{--}=1$. It is also convenient to define a
normalized angular decay distribution
$\widehat{W}(\theta)=W(\theta)/W$ where
$W=\int_{-1}^1W(\theta)d\cos\theta=\sum_{m}H_{mm}$ such that
$\int_{-1}^1\widehat{W}(\theta)d\cos\theta=1$ (later on we denote
$W=\sum_mH_{mm}$ by $H_{U+L}$). Correspondingly we define normalized polarized
decay functions $\hat{H}_{m'm'}=H_{m'm'}/\sum_{m}H_{mm}$.

The distribution Eq.~(\ref{Wtheta}) or its normalized form
$\widehat{W}(\theta)$ is a second order equation in $\cos\theta$, i.e.\ it is
the equation of a parabola with coefficients given by sums of the products
$\rho_{mm}H_{m'm'}$. The parabola is upward bent for
$(1-3\rho_{00})(1-3\hat{H}_{00})>0$ and downward bent for
$(1-3\rho_{00})(1-3\hat{H}_{00})<0$. As a measure of the flatness of the decay
distribution we define a convexity parameter $c_f$ given by the differential
change of slope (or the second derivative) of the decay distribution. From
Eq.~(\ref{Wtheta}) one has
\begin{equation}\label{convex}
c_f=\frac{d^2\widehat{W}}{d(\cos\theta)^2}
  =\frac34(1-3\rho_{00})(1-3\hat{H}_{00})\,.
\end{equation}
As a second global measure we introduce the forward--backward asymmetry of the
decay distribution defined by
\begin{equation}\label{fb}
A_{FB}=\frac{W(F)-W(B)}{W(F)+W(B)}
  =\frac34(\rho_{++}-\rho_{--})(\hat{H}_{++}-\hat{H}_{--})
\end{equation}
where $W(F)=W(0\le\theta\le\pi/2)$ and $W(B)=W(\pi/2\le\theta\le\pi)$.

Of interest is also the location of the extremum of the parabola in
Eq.~(\ref{Wtheta}). The extremum is located at
\begin{equation}\label{extr}
\cos\theta\,\Big|_{\,\rm extr}=-\ \frac{(\rho_{++}-\rho_{--})}{(1-3\rho_{00})}
\ \frac{(\hat{H}_{++}-\hat{H}_{--})}{(1-3\hat{H}_{00})}\,.
\end{equation}
The three measures are not independent for the normalized parabolic decay
distributions $\widehat{W}(\theta)$. They are related by
\begin{equation}
\cos\theta\,\Big|_{\rm\,extr}=-\frac{A_{FB}}{c_f}\,.
\end{equation}
Note that all three measures factor into a production part described by the 
density matrix elements $\rho_{mm}$ and a decay part given in terms of the
polarized decay functions $\hat{H}_{mm}$.

As a well-known illustration consider the cascade decay $t\to b+W^+(\uparrow)$
followed by $W^+(\uparrow)\to f_1\bar{f}_2$. At the Born term level and for
massless fermions (quarks or leptons) the only nonvanishing polarized decay
function in the Standard Model (SM) is $H_{++}$ if one chooses $z'$ to lie
along $\bar{f}_2\in\{\ell^+,\bar q_2\}$. In the normalized form one has
$\hat{H}_{++}=1$. One obtains
\begin{equation}\label{Wtheta1}
\widehat{W}(\theta)=\frac38(1+\cos\theta)^2\,\rho_{++}
  +\frac38(1-\cos\theta)^2\,\rho_{--}+\frac34\sin^2\theta\,\rho_{00}\,.
\end{equation}
The decay distribution~(\ref{Wtheta1}) corresponds to a normalized parabola 
with convexity parameter $c_f=3(1-3\rho_{00})/4$, a maximum at
$\cos\theta\,\big|_{\rm\,extr}=-(\rho_{++}-\rho_{--})/(1-3\rho_{00})$ and a
forward--backward asymmetry of $A_{FB}=\frac34(\rho_{++}-\rho_{--})$.

It is evident that an analysis of an angular decay distribution such as in
Eq.~(\ref{Wtheta1}) can be used to experimentally extract information on the
helicity fractions $\rho_{mm}$ ($m=\pm1,0$) of the $W$ boson. Such an analysis
has been widely applied using the leptonic decay modes of the $W$ boson
$W^+(\uparrow)\to\ell^+\nu_\ell$ (see e.g.\
Refs.~\cite{Aaltonen:2012rz,Peters:2011an}). Clearly the decay
distribution~(\ref{Wtheta1}) will be affected by radiative corrections and
finite mass effects in as much as the three normalized decay structure
functions $\hat{H}_{mm}$ will change from the simple pattern $\hat{H}_{++}=1$;
$H_{00}=H_{--}=0$ on which Eq.~(\ref{Wtheta1}) is based. The modification of
Eq.~(\ref{Wtheta1}) in the quark sector through radiative corrections and
finite mass effects is the
subject of this paper. In this context it is important to note that for
$t\to b+W^+(\uparrow)(\to q_{1}\bar{q}_{2})$ the radiative corrections to 
$t\to b+W^+(\uparrow)$ and $W^+(\uparrow) \to q_{1}\bar{q}_{2}$ factorize at NLO
in QCD (but not in higher orders), i.e.\ there is no NLO cross-talk between
the production and decay processes~\cite{Brandenburg:2002xr}.

Note that for unpolarized $(W,Z)$ decay, when
$\rho_{++}=\rho_{00}=\rho_{--}=1/3$, Eq.~(\ref{Wtheta}) leads to a flat decay
distribution
\begin{equation}
W(\theta)=\frac12(H_{++}+H_{00}+H_{--})\,.
\end{equation}

In the next two sections we shall first write out the diagonal spin density
matrix elements $\rho_{mm}$ of the $(W,Z)$ boson in two prominent sample
production processes and then, at a later stage, proceed to calculate the
relevant set of decay structure functions $H_{mm}$ at $O(\alpha_s)$. The
results are then combined to present analytical and numerical results for the
respective $O(\alpha_s)$ decay distributions. 

%%%%%%%%%%%%%%%%%%%%%%%%%%%%%%%%%%%%%%%%%%%%%%%%%%%%%%%%%%%%%%%%%%%%%%%%%%%%%
\section{Spin density matrix of the $W$ boson\\
  in the decay $t\to b+W^+(\uparrow)$}
%%%%%%%%%%%%%%%%%%%%%%%%%%%%%%%%%%%%%%%%%%%%%%%%%%%%%%%%%%%%%%%%%%%%%%%%%%%%%
Let us start by discussing the spin density matrix of the $W^+$ in the decay
$t\to b+W^+(\uparrow)$. As indicated by the notation, the $W$ boson emerges in
a polarized state in this decay.

The spin density matrix elements of the $W^+$ in $t\to b+W^+$ have been well
studied. We take the $z$ axis to lie along the momentum of the $W^+$ in the
top quark rest frame. At LO one has~\cite{Kane:1991bg}
\begin{eqnarray}
\label{spinden}
\rho_{++}({\it Born\/})&=&0\qquad\qquad\qquad\qquad\quad\to\,\,0.0007\,,\\[9pt]
\rho_{00}({\it Born\/})&=&\frac1{1+2x^2}\ =0.696\qquad\to\,\,\,0.6887\,,
  \nonumber\\[3pt]
\rho_{--}({\it Born\/})&=&\frac{2x^2}{1+2x^2}\ =\ 0.304\qquad\!\to\,\,0.3106\,,
  \nonumber
\end{eqnarray}
where $x=m_W/m_t$. For the numerical values we use the central values of
$m_W=80.399\pm0.025\GeV$ and $m_t=172.0\pm0.9\pm1.3\GeV$ provided by the
Particle Data Group~\cite{Nakamura:2010zzi}. In Eq.~(\ref{spinden}) we have 
also given the NLO QCD results indicated by arrows (cf.\
Refs.~\cite{Fischer:1998gsa,Fischer:2000kx,%
Fischer:2001gp,Do:2002ky}).\footnote{The NNLO corrections to the spin density
matrix elements of the $W^+$ have recently been calculated in
Ref.~\cite{Czarnecki:2010gb}.} The radiative correction to $\rho_{++}$ can be
seen to be very small.
The absolute (relative) corrections to $\rho_{00}$ and $\rho_{--}$ amount to 
$-0.73\%$ ($-1.05\%$) and $0.66\%$ ($2.17\%$).

The spin density matrix elements in Eq.~(\ref{spinden}) are calculated for
unpolarized top decays. If the decaying top quark is polarized, the spin
density matrix elements will depend on the orientation $\theta_P$ and the
degree of polarization $P_t=|\vec{P}_t|$ of the top quark. The dependence is
very simple for the $m_b=0$ Born term case where one has
\begin{equation}\label{rhowpol}
\rho^{P}_{++}=0,\qquad
\rho^{P}_{00}=\rho_{00}\frac{1+P_{t}\cos\theta_P}{D(\theta_P)},\qquad
\rho^{P}_{--}=\rho_{--}\frac{1-P_{t}\cos\theta_P}{D(\theta_P)},
\end{equation}
and where the denominator in Eq.~(\ref{rhowpol}) is given by
\begin{equation} 
D(\theta_{P})=\rho_{00}(1+P_t\cos\theta_P)+\rho_{--}(1-P_t\cos\theta_P).
\end{equation}
It is clear that the relative weight of the two helicity fractions $\rho_{00}$
and $\rho_{--}$ can be changed by appropiately tuning the polarization of the
top quark.

In the general case both helicity states of the polarized top quark are 
involved. This case is somewhat more complicated and has been worked out 
in Refs.~\cite{Fischer:1998gsa,Fischer:2000kx}.

%%%%%%%%%%%%%%%%%%%%%%%%%%%%%%%%%%%%%%%%%%%%%%%%%%%%%%%%%%%%%%%%%%%%%%%%%%%%%
\section{Spin density matrix of the $Z$ boson\\
  in the production process $e^+e^- \to Z(\uparrow)$}
%%%%%%%%%%%%%%%%%%%%%%%%%%%%%%%%%%%%%%%%%%%%%%%%%%%%%%%%%%%%%%%%%%%%%%%%%%%%%
Next we discuss the polarization of the $Z$ boson produced in $e^+e^-$
annihilation where the polarization density matrix of the $Z$ boson is purely
transverse in the $e^+e^-$ system. We take the $z$ axis to lie along the $e^-$
beam ($z\parallel e^-$). The Born term matrix element is given by 
(see e.g.\ Ref.~\cite{Barger:1987nn})
\begin{equation}
{\cal M}(m)=-i\frac{g_Z}4\bar{v}(e^+)\left(v_e\gamma_{\mu}
-a_e\gamma_{\mu}\gamma_5 \right)u(e^-)\epsilon^{\ast \mu}(m)
\end{equation}
($m=0,\pm$) where, in the SM, one has
\begin{equation}
v_\ell=-1+4\sin^2\Theta_W\qquad 
a_\ell=-1 \qquad {\rm for}\quad \ell=\,e,\mu,\tau,
\end{equation}
and where $\Theta_W$ is the Weinberg angle. In our numerical analysis we take
$\sin^2\Theta_W=0.231$~\cite{Nakamura:2010zzi}. One can then work out the
normalized diagonal spin density matrix elements of the $Z$ boson in
$e^+e^- \to Z(\uparrow)$ which are given by
\begin{equation}\label{rhoz}
\rho_{++}=\frac{(v_e-a_e)^2}{2(v_e^2+a_e^2)}=0.4244,\qquad
\rho_{00}=0,\qquad
\rho_{--}=\frac{(v_e+a_e)^2}{2(v_e^2+a_e^2)}=0.5756.
\end{equation}
Note that the two transverse density matrix elements in Eq.~(\ref{rhoz}) are
approximately equal due to the smallness of the vector current coupling
constant $v_e=-0.076$.

Similar to the case $t \to b+W^+$, the spin density matrix elements of the 
$Z$ boson can be tuned by polarizing the $e^+e^-$ beams. Take, for example,
longitudinally polarized beams and denote the longitudinal polarization of the
$e^{\mp}$ beams by $h^\mp$, where the polarization is measured w.r.t.\ the 
momenta of the $e^\mp$ beams.
One then has
\begin{equation}
\label{rhozpol}
\rho^P_{++}=\rho_{++}\frac{(1+h^-)(1-h^+)}{D(h^-,h^+)},\qquad
\rho^P_{00}=0,\qquad
\rho^P_{--}=\rho_{--}\frac{(1-h^-)(1+h^+)}{D(h^-,h^+)},
\end{equation}
where
\begin{equation}
D(h^-,h^+)=\rho_{++}(1+h^-)(1-h^+)+\rho_{--}(1-h^-)(1+h^+).
\end{equation}
For example, with a 100\% longitudinally polarized electron beam one obtains
$\rho^P_{--}=1$ for $h^-=-1$ and $\rho^P_{++}=1$ for $h^-=+1$.

%%%%%%%%%%%%%%%%%%%%%%%%%%%%%%%%%%%%%%%%%%%%%%%%%%%%%%%%%%%%%%%%%%%%%%%%%%%%%
\section{Polarized $W^\pm$ decays into massive quark pairs}
%%%%%%%%%%%%%%%%%%%%%%%%%%%%%%%%%%%%%%%%%%%%%%%%%%%%%%%%%%%%%%%%%%%%%%%%%%%%%
We first treat the case $W^+\to q_1\bar{q}_2$.
The LO Born term amplitude is given by (see e.g. \cite{Barger:1987nn})
\begin{equation}
\label{loampl}
{\cal M}_W(m)
  =-i\frac{g_W}{\sqrt2}V_{q_1q_2}\,\,\bar u_1(p_1)\gamma^\mu\frac{1-\gamma_5}2
  v_2(p_2)\,\epsilon^W_\mu(m),
\end{equation}
where $g_W=e/\sin\Theta_W$ is the electroweak coupling constant and the
$V_{q_1q_2}$ are Kobayashi--Maskawa matrix elements. Let us define a reduced
matrix element $\widetilde{\cal M}(m)$ by splitting off the coupling factors
and the factor $1/2$ from the chiral projector such that
\begin{equation}\label{loamp2}
\widetilde{\cal M}_W(m)
  =\bar u_1(p_1)\gamma^\mu(1-\gamma_5)v_2(p_2)\,\epsilon^W_\mu(m).
\end{equation}
The LO polarized decay functions $H_{mm}$ are then obtained from
\begin{equation}\label{helsf}
H_{mm}\ =\ N_c\ \sum_{\rm quark\, spins}
  \widetilde{\cal M}_W(m)\,\widetilde{\cal M}^\dagger_W(m).
\end{equation}

Our aim is to calculate the NLO polarized decay functions $H_{mm}$ of a $W^+$
boson decaying into a heavy quark pair where the $W^+$ has definite spin
quantum numbers $m=\pm,0$. We first discuss a coordinate system where the $z$
axis lies along the quark momentum $(z\parallel q_1)$ (system I). The Born
term and the $\alpha_s$ contributions to the polarized decay functions
$H_{\pm\pm}$ and $H_{00}$ and the unpolarized total decay function
$H_{U+L}=H_{++}+H_{00} +H_{--}$ are expanded up to the second order in the
quark mass ratios $\sqrt{\mu_{1,2}}=m_{1,2}/m_W$ where we make use of the
unexpanded analytical results given in Ref.~\cite{gkt12}. One obtains
\begin{eqnarray}\label{alphas}
H^{\rm I}_{++}&=&8N_cq^2\bigg[\,0+\ldots\nonumber\\&&\qquad\quad\,\,
  +\frac{\alpha_s}{6\pi}\Big(1+(\pi^2-16)\sqrt{\mu_1}+(5+2\pi^2/3)\,\mu_1+\mu_2
  -2\mu_1\ln\mu_1+\ldots\Big)\bigg],\qquad\\
H^{\rm I}_{00}&=&8N_cq^2\bigg[\,0+\frac{(\mu_1+\mu_2)}2+\ldots
  \nonumber\\&&\qquad\quad\,\,
  +\frac{\alpha_s}{6\pi}\Big(4-2\pi^2\sqrt{\mu_1}-(39+8\pi^2/3)\,\mu_1+\mu_2
  -30\mu_1\ln\mu_1\nonumber\\&&\qquad\qquad\qquad
  +6\mu_2\ln\mu_2- 2\mu_1\ln^2\mu_1+\ldots\Big)\bigg],\\
H^{\rm I}_{--}&=&8N_cq^2\bigg[\,1-\mu_1-\mu_2+\ldots
  \nonumber\\&&\qquad\quad\,\,
  +\frac{\alpha_s}{6\pi}\Big(1+(\pi^2+16)\sqrt{\mu_1}+(49+2\pi^2)\,\mu_1
  +13\mu_2+14\mu_1\ln\mu_1\nonumber\\&&\qquad\qquad\qquad
  -24\mu_2\ln\mu_2+ 2\mu_1\ln^2\mu_1+\ldots\Big)\bigg].\label{alphasm}
\end{eqnarray}
We truncate the mass expansion at $O(\mu_i)$ since third order quark mass
effects can be expected to be quite small judging from the fact that
$\mu_c^{3/2}=(1.5/80.399)^3=6.49\cdot10^{-6}$ and
$\mu_b^{3/2}=(4.8/80.399)^3=0.21\cdot10^{-3}$.

For the sum of the three polarized decay functions $\sum_{m}H_{mm}:=H_{U+L}$ one
has
\begin{eqnarray}
\label{alphassum}
H_{U+L}&=&H^{\rm I}_{U+L}
  \ =\ H^{\rm I}_{++}+H^{\rm I}_{00}+H^{\rm I}_{--}\nonumber\\
  &=&8N_cq^2\bigg[\,1-\frac{(\mu_1+\mu_2)}2\,+\ldots
  \nonumber\\&&\qquad\quad \,\,
  +\frac{\alpha_s}{6\pi}\Big(6+15\mu_1+15\mu_2
  -18\mu_1\ln\mu_1-18\mu_2\ln\mu_2+\ldots\Big)\bigg].\qquad
\end{eqnarray}
Note that the sum of the polarized decay functions is independent of the
choice of the $z$ axis as is indicated in Eq.~(\ref{alphassum}).

The mass corrections set in quadratically in the LO polarized decay functions
and also in the radiatively corrected unpolarized decay function $H_{U+L}$. In
contrast to this, the mass corrections to the radiatively corrected polarized
decay functions $H_{\pm\pm}$ and $H_{00}$ set in linearly. Surprisingly, some
of the linear mass corrections carry rather large coefficients such as the
coefficient $(\pi^2+16)=25.87$ multiplying the linear mass term $\sqrt{\mu_1}$
in $H_{--}^{\rm I}$ in Eq.~(\ref{alphasm}). Contrary to naive expectations one
therefore needs to keep finite mass effects in the radiatively corrected
polarized decay functions for massive quark pair production even at the
relatively large mass scale of the $W$ mass. Note that the radiatively
corrected unpolarized decay function $H_{U+L}$ is symmetric in the quark
masses whereas the polarized decay functions show a large quark mass asymmetry
at NLO.

In order to make contact with the unpolarized decay rate
$\Gamma(W^+\to q_1\bar{q}_2)$, we define polarized decay rates
\begin{equation}
\Gamma_{mm}=\frac{|\vec{p}|}{64\pi m_W^2}g_W^2|V_{q_1q_2}|^2\,H_{mm}
\end{equation}
($|\vec{p}|=\frac{m_W}2(1+\mu_1^2+\mu_2^2-2\mu_1-2\mu_2-2\mu_1\mu_2)^{1/2}$)
which, for the unpolarized rate, gives
\begin{equation}
\Gamma(W^+\to q_1\bar{q}_2)=\frac13(\Gamma_{++}+\Gamma_{00}+\Gamma_{--})=
  \frac{|\vec{p}|}{192\pi m_W^2}g_W^2|V_{q_1q_2}|^2\,H_{U+L}.
\end{equation}

The quark mass corrections will be most important for the decay
$W^+\to c\bar{b}$. For the quark masses we take $m_c=1.5\GeV$ and
$m_b=4.8\GeV$. For the $W^+\to c\bar{b}$ unpolarized decay function 
$H_{U+L}$ one obtains
\begin{equation}\label{uplusl1}
H_{U+L}=8N_cq^2\Big[\,0.998+\,\ldots
  +\frac{\alpha_s}{6\pi}\big(6+0.470+\ldots\big)\Big].
\end{equation}
In order to highlight the numerical importance of quark mass effects we have
separately listed the leading and the $O(\mu_i)$ quark mass effects in the
NLO terms in Eq.~(\ref{uplusl1}). Even though the mass corrections to the sum
of the polarized decay functions set in only quadratically, the mass effects
in the NLO corrections can be seen to amount to a non-negligible $O(8\%)$ 
where the largest contribution comes from the bottom quark term
$-18\mu_2\ln\mu_2=0.0071$. Using $\alpha_s(m_W^2)=0.117$ one finds an overall
increase of the zero mass Born term decay rate by $3.8\%$, i.e.\ one has
\begin{equation}\label{ratecor}
\Gamma(\,{\rm NLO};O(\mu_{i}))=1.038\,\,\cdot\Gamma(\,{\rm Born};\mu_{i}=0)\,,
\end{equation}
where the bulk of the increase comes from the NLO zero mass term.

 For the $W^+\to c\bar{b}$ polarized decay functions one obtains
\begin{eqnarray}\label{masscor0}
H^{\rm I}_{++}&=&8N_cq^2\,\big[\,0+\,\ldots\phantom{.002}
  +\frac{\alpha_s}{6\pi}\,\big(1-0.101+\ldots\big)\,\big],\nonumber\\
H^{\rm I}_{00}&=&8N_cq^2\,\big[\,0.002+\,\ldots
  +\frac{\alpha_s}{6\pi}\,\big(4-0.469+\ldots\big)\,\big],\nonumber\\
H^{\rm I}_{--}&=&8N_cq^2\,\big[\,0.996+\,\ldots
  +\frac{\alpha_s}{6\pi}\,\big(1+1.040+\ldots\big)\,\big].
\end{eqnarray}
The mass corrections to the NLO polarized decay functions can be seen to be
large. The largest mass correction occurs for $H^{\rm I}_{--}$ which is the only
polarized decay function which is nonzero at the $m_q=0$ Born term level. The
mass correction to the leading NLO contribution in $H^{\rm I}_{--}$ is of
$O(100\%)$.

The large NLO mass correction to $H^{\rm I}_{--}$ does not, however, feed
through to the normalized decay distribution which is governed by the
normalized polarized decay functions $\hat{H}^{I}_{mm}=H^{I}_{mm}/H_{U+L}$.
This can be appreciated by writing out the polarized decay functions in a
generic notation where the (small) mass corrections to the Born term
contributions are neglected. One has
\begin{eqnarray}\label{Hsymbol}
H_{++}&=&8N_{c}q^{2}(0+\frac{\alpha_{s}}{6\pi}(1+\mu_{++}))\nonumber \\
H_{00}&=&8N_{c}q^{2}(0+\frac{\alpha_{s}}{6\pi}(4+\mu_{00}))\nonumber \\
H_{--}&=&8N_{c}q^{2}(1+\frac{\alpha_{s}}{6\pi}(1+\mu_{--}))
\end{eqnarray}
where the $\mu_{mm}$ denote the respective NLO finite mass corrections. One 
then expands the normalized polarized decay functions $\hat H_{mm}$ in the 
strong coupling constant $\alpha_s$. In this approximation one has
\begin{eqnarray}\label{approxh}
\hat H_{++}&=&0+\frac{\alpha_{s}}{6\pi}(1+\mu_{++})\nonumber \\
\hat H_{00}&=&0+\frac{\alpha_{s}}{6\pi}(4+\mu_{00})\nonumber \\
\hat H_{--}&=&1+\frac{\alpha_{s}}{6\pi}(-5-\mu_{++}-\mu_{00}).
\end{eqnarray}
It is apparent that the NLO finite mass corrections to the normalized
polarized decay functions $\hat H_{mm}$ are solely determined by the $O(10\%)$
mass corrections $\mu_{++}$ and $\mu_{00}$. Since $\mu_{++}$ and $\mu_{00}$ are
negative, the NLO mass corrections are destructive. Numerically the NLO mass
corrections amount to only $O(-10\%)$ of the leading NLO mass term (see
Eq.~(\ref{masscor0})). The $O(100\%)$ NLO mass correction to $H_{--}$ drops out
when normalizing the polarized decay functions.   

We are now in the position to write down numerical results for the angular
decay distribution (\ref{Wtheta}). We take into account $O(\alpha_s)$ results
both for the density matrix elements $\rho_{mm}$ and the decay structure
functions $H_{mm}$. At $O(\alpha_s)$ one thus has to take the sum
$\rho_{mm}({\it Born\/})H_{m'm'}(\alpha_s)
+\rho_{mm}(\alpha_s)H_{m'm'}({\it Born\/})$. As mentioned before, there is no
$O(\alpha_s)$ cross-talk between the production and the decay process because
the intermediate gauge boson is colour neutral~\cite{Brandenburg:2002xr}. 
As before we concentrate on the decay $W^+\to c\bar{b}$.
For system~I one finds ($\alpha_s(m_W^2)=0.117$)
\begin{equation}\label{angdist1}
\widehat{W}^{\rm I}(\theta)
=
\frac38(1+\cos\theta)^2\Bigg\{ 
\begin{tabular}{lr}
0.305\\[-2.0ex]
0.318\\[-2.0ex]
0.318
\end{tabular}
\Bigg\}
 +\frac38(1-\cos\theta)^2\Bigg\{ 
\begin{tabular}{lr}
0.001\\[-2.0ex]
0.019\\[-2.0ex]
0.018
\end{tabular}
 \Bigg\}
+\frac34\sin^2\theta\,\Bigg\{ 
\begin{tabular}{lr}
0.694\\[-2.0ex]
0.663\\[-2.0ex]
0.664
\end{tabular}
\Bigg\}\qquad
\end{equation}
For the sake of comparison we have listed three numerical values each for
the angular coefficients. The top entry is for $({\rm Born};\,\mu_{i}\neq 0)$,
the middle entry is for $(O(\alpha_s);\,\mu_{i}=0)$, and the bottom entry is
for $(O(\alpha_s);\,\mu_{i}\neq 0)$. The same three-tiered notation will be 
used in subsequent formulas and in Table~\ref{tab1}.

\begin{figure}[t]
\epsfig{figure=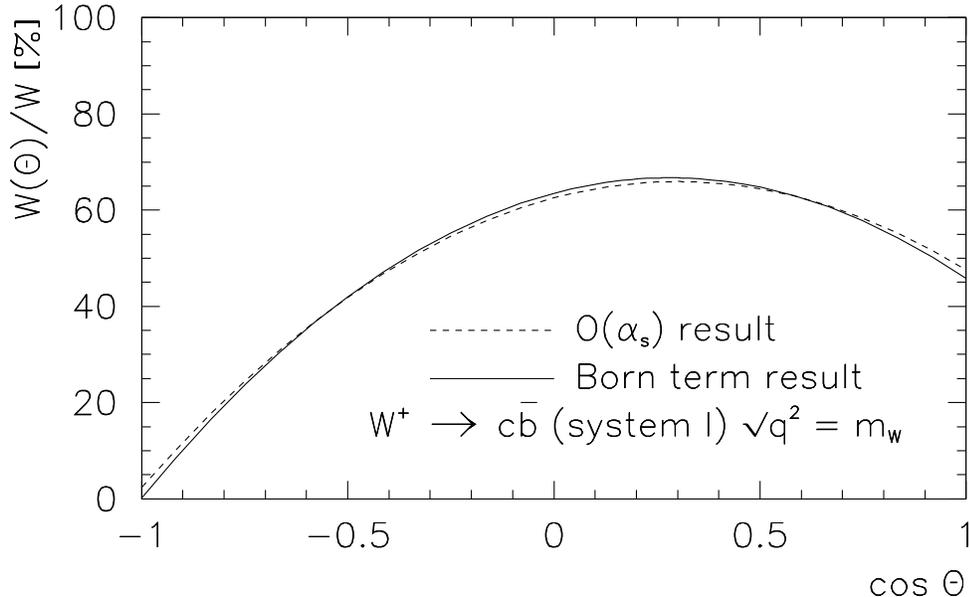, scale=0.8}
\caption{\label{wfunw}Normalized angular decay distribution
$\widehat{W}(\theta)=W(\theta)/W$ for $W^+\to c\bar b$ at LO (full line) and
NLO (dotted line) in system I including $O(m_{c,b}^2)$ finite quark mass
contributions.}
\end{figure}

Eq.~(\ref{angdist1}) describes a downward bent parabola with unit area and,
as Fig.~\ref{wfunw} shows, a maximum slightly displaced to the right of
$\cos\theta=0$. By comparing the respective numbers in the normalized decay
distribution $\widehat{W}(\theta)$ in Eq.~(\ref{angdist1}), NLO quark mass
effects can be seen to be almost negligible even if they are important for the
polarized decay functions. More important are the radiative corrections
which lead to a $2.0\%$ enhancement (on an absolute scale) of the normalized
decay  distribution at the forward point ($\cos\theta=1$) and a $2.7\%$
enhancement at the backward point ($\cos\theta=-1$). Midways at $\cos\theta=0$
one finds a $1.2\%$ depletion of the decay distribution. This is illustrated
in Fig.~\ref{wfunw} where we compare the $O(\mu_i)$ normalized angular decay
distributions for the Born term case and the $O(\alpha_s)$ case. The net
effect of the radiative corrections is to make the normalized angular decay
distribution flatter with little dependence on quark mass effects. 

In Table~\ref{tab1} we present our numerical results for the three global
parameters $c_f$, $A_{FB}$ and $\cos\theta\,\Big|_{\,\rm extr}$ that characterize
the decay distribution.  In Table~1 we use the same three-tiered notation as
in Eq.~(\ref{angdist1}). The numerical results for system~I ($z\,\parallel\,c$)
are listed in column~1. We emphasize that the inverse of the denominators
factors occurring in the calculation of the global measures in
Eqs.~(\ref{convex}--\ref{extr}) have been left unexpanded in $\alpha_s$ when
calculating the entries in Table~1. The initial and final state QCD
corrections can be seen to reduce the convexity parameter $c_f$ and the
forward-backward asymmetry by $8.37\,\%$ and $1.25\,\%$, respectively. The
maximum of the decay distribution is shifted to the right by $7.7\,\%$. By
comparing the numbers in tier~3 with those in tier~2 one can see that the bulk
of these shifts come from the NLO zero mass corrections. NLO finite mass
effects are small and tend to slightly reduce the NLO zero mass corrections. 

The largest NLO corrections come from the final state corrections. One can
therefore obtain a rough understanding of the numbers in Table~1 by neglecting
the initial state corrections and the (small) finite mass effects in the Born
term contributions. Expanding the relevant contributions to $O(\alpha_s)$ one
obtains
\begin{eqnarray}
c^{\rm I}_f&=&\frac34(1-3\rho_{00})\Big(1-\frac{\alpha_s}{6\pi}(12-1.41)\Big)
  \ =\ -0.762\,,\label{rough1}\\
A^{\rm I}_{FB}&=&-\frac34(\rho_{++}-\rho_{--})\Big(1-\frac{\alpha_s}{6\pi}
  (6-0.671)\Big)\ =\ 0.221\,,\label{rough2}\\
\cos\theta\,\Big|^{\rm I}_{\,\rm max}&=&\frac{\rho_{++}-\rho_{--}}{1-3\rho_{00}}\,
  \Big(1+\frac{\alpha_s}{6\pi}(6-0.736)\Big)=0.289\,.\label{rough3}
\end{eqnarray}
Looking at the different contributions in Eq.~(\ref{rough1}--\ref{rough3}) one
can understand the main features of the numerical results listed in Table~1
that were already discussed above. 

In Eqs.~(\ref{rough1}--\ref{rough3}) we also quote approximate numbers which
are calculated from the finite mass corrections $\mu_{++}$ and $\mu_{--}$ listed
in Eq.~(\ref{masscor0}) and the Born term values for the spin density elements
$\rho_{mm}$ listed in Eq.~(\ref{spinden}). The approximate numbers listed in
Eqs.~(\ref{rough1}--\ref{rough3}) can be seen to deviate by small amounts from
the exact numbers listed in Table~1 where the largest deviation occurs for
$A^{\rm I}_{FB}$ and $\cos\theta\,\Big|^{\rm I}_{\,\rm max}$. 
\begin{table}[ht]\begin{center}
\begin{tabular}{|l||c|c|c|c|}\hline
&\multicolumn2{|c|}{$W^+\to c\bar b$}&$Z\to c\bar c$&$Z\to b\bar b$\\
&system I ($z\,\parallel\,c$)&system II ($z\,\parallel\,\bar b$)
&system I ($z\,\parallel\,c$)&system I ($z\,\parallel\,b$)\\\hline
&$-0.811$&$-0.811$&$+0.750$&$+0.746$\\[-2.0ex]$c_f$
&$-0.741$&$-0.741$&$+0.697$&$+0.697$\\[-2.0ex]
&$-0.743$&$-0.752$&$+0.701$&$+0.706$\\\hline
&$+0.228$&$-0.228$&$+0.076$&$+0.106$\\[-2.0ex]$A_{FB}$
&$+0.224$&$-0.224$&$+0.073$&$+0.102$\\[-2.0ex]
&$+0.225$&$-0.226$&$+0.073$&$+0.103$\\\hline
&$+0.281$&$-0.281$&$-0.101$&$-0.142$\\[-2.0ex]
$\cos\theta\,\Big|_{\rm\,extr}$
&$+0.304$&$-0.304$&$-0.105$&$-0.147$\\[-2.0ex]
&$+0.302$&$-0.301$&$-0.105$&$-0.146$\\\hline
\end{tabular}
\caption{\label{tab1}The parameters $c_f$, $A_{FB}$ and
$\cos\theta\,\Big|_{\, \rm extr}$ characterizing the normalized polar angle decay
distribution of the cascade decays $t\to b+W^+(\uparrow)(\to c\bar{b})$ and
and the production/decay process
$e^+e^-\to Z(\uparrow)\to b\bar{b},c\bar{c}$. We use a three-tiered notation
where the top entries are for $({\rm Born};\,\mu_{i}\neq 0)$,
the middle entry is for $(O(\alpha_s);\,\mu_{i}=0)$, and the bottom entry is
for $(O(\alpha_s);\,\mu_{i}\neq 0)$.}
\end{center}\end{table}

Summarizing our results in system~I one finds that NLO and NLO quark mass
effects are important for the polarized decay functions. The NLO quark mass
effects, however, do not feed through to the angular decay distribution which
is mainly affected by the leading mass term in the final state radiative
correction. One concludes that quark mass effects are quite small for the
radiatively corrected normalized angular decay distribution even if they are
important for the radiatively corrected decay functions.  

When the polar angle $\theta$ is measured w.r.t.\ the antiquark direction
$(z\parallel\bar{q})$ (system~II), the relevant expressions for the helicity
structure functions can be obtained from those in system~I by the exchange
$\mu_1\leftrightarrow\mu_2$, and $H_{\pm\pm}\leftrightarrow H_{\mp\mp}$ and
$H_{00}\leftrightarrow H_{00}$. One has
\begin{equation}
\label{1and2}
H^{\rm II}_{\mp\mp}(\mu_1,\mu_2)=H^{\rm I}_{\pm\pm}(\mu_2,\mu_1),\qquad
H^{\rm II}_{00}(\mu_1,\mu_2)=H^{\rm I}_{00}(\mu_2,\mu_1).
\end{equation}
Because of the mass asymmetry of the polarized decay functions and because one
is exchanging $m_c\leftrightarrow m_b$, the quark mass effects are
more pronounced in system II. In fact, one now has
\begin{eqnarray}
H^{\rm II}_{++}&=&8N_cq^2\,\big[\,0.996+\,\ldots
  +\frac{\alpha_s}{6\pi}\,\big(1+1.806+\ldots\big)\,\big],\nonumber\\
H^{\rm II}_{00}&=&8N_cq^2\,\big[\,0.002+\,\ldots
  +\frac{\alpha_s}{6\pi}\,\big(4-1.051+\,\ldots\big)\,\big],\nonumber\\
H^{\rm II}_{--}&=&8N_cq^2\,\big[\,0+\,\ldots\phantom{.002}
  +\frac{\alpha_s}{6\pi}\,\big(1-0.284+\,\ldots\big)\,\big].
\end{eqnarray}
The NLO quark mass corrections in system~II can be seen to be approximately
two-and-a half times larger than those in system~I. The largest mass
correction now occurs for $H^{\rm II}_{++}$.

Numerically one obtains
\begin{equation}\label{angdist2}
\widehat{W}^{\rm II}(\theta)
=\frac38(1+\cos\theta)^2\Bigg\{ 
\begin{tabular}{lr}
0.001\\[-2.0ex]
0.019\\[-2.0ex]
0.016
\end{tabular}
\Bigg\}
+\frac38(1-\cos\theta)^2\Bigg\{ 
\begin{tabular}{lr}
0.305\\[-2.0ex]
0.318\\[-2.0ex]
0.317
\end{tabular} 
\Bigg\}
+\frac34\sin^2\theta\,\Bigg\{ 
\begin{tabular}{lr}
0.694\\[-2.0ex]
0.663\\[-2.0ex]
0.667
\end{tabular} 
\Bigg\}.\qquad
\end{equation}
Again one observes that the large quark mass effects seen in the polarized
decay functions do not feed through to the angular decay distribution.
Of course, the reason is the same as explained after Eq.~(\ref{approxh}).
Except for the slightly enhanced NLO quark mass effects, the distribution 
(\ref{angdist2}) is just a reflection of Eq.~(\ref{angdist1}) at the line 
$\cos\theta=0$, i.e. $A^{\rm I}_{FB}\sim -A^{\rm II}_{FB}$. 

Larger quark mass effects can also be seen in the convexity parameter 
$c^{\rm II}_f$, in the forward-backward asymmetry $A^{\rm II}_{FB}$ and in the
position of the maximum $\cos\theta\,\Big|^{\rm II}_{\,\rm max}$ (see Table~1). The
larger size of the NLO finite mass corrections to the global measures can be
inferred by listing approximate formulas similar to those in
Eqs.~(\ref{rough1}--\ref{rough3}). One has
\begin{eqnarray}
c^{\rm II}_f&=&\frac34 (1-3\rho_{00})\Big(1-\frac{\alpha_s}{6\pi}(12-3.153)\Big)
  \ =\ -0.771\,,\label{rough4}\\
A^{\rm II}_{FB}&=&-\frac34(\rho_{++}-\rho_{--})\Big(1-\frac{\alpha_s}{6\pi}
  (6-1.619)\Big)\ =\ 0.222\,,\label{rough5}\\
\cos\theta\,\Big|^{\rm II}_{\,\rm max}&=&\frac{\rho_{++}-\rho_{--}}{1-3\rho_{00}}\,
  \Big(1+\frac{\alpha_s}{6\pi}(6-1.534)\Big)=0.287\,.\label{rough6}
\end{eqnarray}
One can see that the NLO finite mass effects make up $\sim 25\%$ of the
leading NLO terms as compared to the $\sim 10\%$ in system I (see
Eqs.~(\ref{rough1}--\ref{rough3})). Again the approximate numbers listed in
Eqs.~(\ref{rough4}--\ref{rough6}) can be seen to only deviate by small amounts
from the exact numbers listed in Table~1 (column 2; tier 3). Again one
concludes that, even though quark mass effects are larger in system II, the
bulk of the radiative corrections still come from the NLO leading terms
$\propto 12\alpha_s/(6\pi)$ and $\propto 6\alpha_s/(6\pi)$.

Up to this point we have only considered the decay $W^+\to q_1\bar{q}_2$.
The charge conjugated decay
$W^-\to \bar{q}_1q_2$ is related to 
$W^+\to q_1\bar{q}_2$ by $CP$-invariance. The corresponding helicity
structure functions $H_{mm}(W^-\to \bar{q}_1q_2)$ can be obtained via
the relations
\begin{equation}\label{wminus}
H_{mm}(W^-\to\bar{q}_1q_2;\mu_1,\mu_2;z\parallel q_2)=
H_{mm}(W^+\to q_1\bar{q}_2;\mu_2,\mu_1;z\parallel q_1),
\end{equation}
where the $z$ axis for the decay $W^-\to\bar{q}_1q_2$ lies along the
quark direction $(z\parallel q_2)$.

%%%%%%%%%%%%%%%%%%%%%%%%%%%%%%%%%%%%%%%%%%%%%%%%%%%%%%%%%%%%%%%%%%%%%%%%%%%%%%
\section{Polarized $Z$ decays into massive quark pairs}
%%%%%%%%%%%%%%%%%%%%%%%%%%%%%%%%%%%%%%%%%%%%%%%%%%%%%%%%%%%%%%%%%%%%%%%%%%%%%%
In the SM the Born term matrix element for the decay $Z(m)\to q\bar{q}$ 
with spin quantum numbers $m$ is given by (see e.g.\ Ref.~\cite{Barger:1987nn})
\begin{equation}
{\cal M}_Z(m)=-i\frac{g_Z}4\bar{u}(q)\left(v_f\,\gamma_\mu
  -a_f\gamma_\mu\gamma_5\right)v(\bar{q}) \epsilon^\mu(m)
\end{equation}
where $g_Z^2=8G_FM_Z^2/\sqrt2$, and where, in the SM, one has
\begin{eqnarray}
v_f&=&1-\frac83\sin^2\Theta_W,\qquad a_f=1\qquad{\rm for}\quad u,c,t,\\
v_f&=&-1+\frac43\sin^2\Theta_W,\qquad a_f=-1\qquad{\rm for}\quad d,s,b.
\end{eqnarray}
Similar to Eq.~(\ref{loamp2}) we define reduced amplitudes by writing
\begin{equation}
\widetilde{\cal M}_Z(m)=\bar{u}(q)\left(v_f\,\gamma_\mu
  -a_f\gamma_\mu\gamma_5\right)v(\bar{q})\epsilon^\mu(m).
\end{equation}
As in Eq.~(\ref{helsf}), the LO polarized decay functions $H_{mm}$ are
calculated according to 
\begin{equation}\label{helsf1}
H_{mm}=\ N_c\ \sum_{\rm quark\, spins}
  \widetilde{\cal M}_Z(m)\,\widetilde{\cal M}_Z^\dagger(m).
\end{equation}

Compared to the charged current case the relative weights of the vector ($V$)
and axial vector current ($A$) contributions are no longer simple and it is
more convenient to switch to a notation in terms of the $VV$, $AA$ and $VA=AV$
contributions. Again we make use of the analytical results in
Ref.~\cite{gkt12} (or those in Ref.~\cite{Groote:2010zf}) which we expand up
to $O(\mu)$. One obtains 
\begin{eqnarray}\label{vvaa}
H_{U}^{VV/AA}&=&4N_cq^2\bigg[1-2\mu\pm 2\mu\,\ldots
  +\frac{\alpha_s}{6\pi}\Big(2+2\pi^2\sqrt\mu+(68+\frac{8\pi^2}3)\mu
  \nonumber\\&&\qquad\quad\,\,
  -12\mu\ln\mu+2\mu\ln^2\mu\pm 24\mu\,\,(1+\ln\mu)\,\ldots\Big)\bigg],
  \nonumber\\
H_{F}^{VA/AV}&=&4N_cq^2\bigg[1-2\mu\,\ldots+\frac{\alpha_s}{6\pi}
  \Big(32\sqrt\mu+(56+\frac{4\pi^2}3)\mu-8\mu\ln\mu
  \nonumber\\&&\qquad\quad\,\,
  +2\mu\ln^2\mu\,\ldots\Big)\bigg],\nonumber\\
H_{L}^{VV/AA}&=&4N_cq^2\bigg[\mu\pm \mu\,\ldots+\frac{\alpha_s}{6\pi}\Big(4
  -2\pi^2\sqrt\mu-(38+\frac{8\pi^2}3)\mu\nonumber\\&&\qquad\quad\,\,
  -24\mu\ln\mu-2\mu\ln^2\mu\pm 6\mu\,(5+2\ln\mu)\,\ldots\Big)\bigg].
\end{eqnarray}
Note that, in the zero mass limit, there are no $\alpha_s$ corrections to the
parity-violating structure function $H_{F}^{VA}$, as noted before in
Ref.~\cite{Jersak:1979uv,Korner:1985dt}.

As in system~I of the charged current case we evaluate the polarized decay
functions $H_{\pm\pm}$ and $H_{00}$ in a system where the quark lies along the
$z$ direction $(z\parallel q)$. One has
\begin{eqnarray}\label{helz}
H^{\rm I}_{\pm\pm}&=&\frac12\left(v^2_f H_{U}^{VV}
  +a^2_f H_{U}^{AA}\mp 2v_fa_fH_{F}^{VA}\right),\nonumber\\
H^{\rm I}_{00}&=&v_f^2H_{L}^{VV}+a_f^2H_{L}^{AA}.
\end{eqnarray}
As a check on Eq.~(\ref{helz}) one can set $v_f=a_f=1$ or $v_f=a_f=-1$ and one
will then recover the charged current results Eq.~(\ref{alphas}) (system~I)
with $\mu_1=\mu_2=\mu$. When the $z$ axis is taken to be along the antiquark
direction (system II; $z\parallel\bar{q}$), there will be no change for
$H_{00}$ but one needs to exchange $H_{\pm\pm} \leftrightarrow H_{\mp\mp}$ (or,
$H_{F}^{VA}\leftrightarrow-H_{F}^{VA}$) in Eq.~(\ref{helz}).

For the sum of the three polarized decay functions one obtains
\begin{eqnarray}\label{upluslbb}
H_{U+L}&=&H^{\rm I}_{U+L}\ =\ H^{\rm I}_{++}+H^{\rm I}_{00}+H^{\rm I}_{--}
  \ =\ v_f^2\,H^{VV}_{U+L}+a_f^2\,H^{AA}_{U+L}\nonumber\\
  &=&8N_cq^2\bigg\{v_f^2\Big[\frac12+\mu\,\ldots
  +\frac{\alpha_s}{6\pi}\left(3+42\mu+\ldots\right)\Big] \nonumber\\
  &&\phantom{8N_cq^2\bigg\{}+a_f^2\Big[\frac12-2\mu\,\ldots
  +\frac{\alpha_s}{6\pi}\left(3-12\mu-36\mu\ln\mu+\ldots\right)\Big]\bigg\}.
\end{eqnarray}
In order to make contact with the total decay rate $\Gamma(Z\to q\bar{q})$, we
define polarized decay rates by 
\begin{equation}
\Gamma_{mm}=\frac{G_F|\vec{p}|}{16\pi\sqrt2}H_{mm}
\end{equation}
where $|\vec{p}|=\frac{m_Z}2\sqrt{1-4\mu}$. For the unpolarized decay rate one
then obtains
\begin{equation}
\Gamma(Z\to q\bar{q})=\frac13(\Gamma_{++}+\Gamma_{00}+\Gamma_{--}) 
  =\frac{G_F|\vec{p}|}{48\pi\sqrt2}H_{U+L}.
\end{equation}

Returning to Eq.~(\ref{vvaa}), we write down our numerical results for the
decay $Z\to b\bar{b}$. In system~I one has
\begin{eqnarray}
H_U^{VV}\,(b\bar b)&=&4N_cq^2\,\big[\,1\phantom{.000}
  +\frac{\alpha_s}{6\pi}\,\big(2+1.363+\ldots\big)\,\big],\nonumber\\
H_U^{AA}\,(b\bar b)&=&4N_cq^2\,\big[\,0.989
  +\frac{\alpha_s}{6\pi}\,\big(2+2.013+\ldots\big)\,\big],\nonumber\\
H_F^{VA}\,(b\bar b)&=&4N_cq^2\,\big[\,0.994
  +\frac{\alpha_s}{6\pi}\,\big(0+2.199+\ldots\big)\,\big],\nonumber\\
H_L^{VV}\,(b\bar b)&=&4N_cq^2\,\big[\,0.006
  +\frac{\alpha_s}{6\pi}\,\big(4-1.131+\ldots\big)\,\big],\nonumber\\
H_L^{AA}\,(b\bar b)&=&4N_cq^2\,\big[\,0\phantom{.000}
  +\frac{\alpha_s}{6\pi}\,\big(4-0.905+\ldots\big)\,\big],
\end{eqnarray}
or, using Eq.~(\ref{helz})
\begin{eqnarray}
H^{\rm I}_{++}(b\bar b)&=&4N_cq^2\,\big[\,0.046
  +\frac{\alpha_s}{6\pi}\,\big(1.479-0.189 +\ldots\big)\,\big],\nonumber\\
H^{\rm I}_{00}(b\bar b)&=&4N_cq^2\,\big[\,0.003
+\frac{\alpha_s}{6\pi}\,\big(5.916-1.447+\ldots\big)\,\big],\nonumber\\
H^{\rm I}_{--}(b\bar b)&=&4N_cq^2\,\big[\,1.422
+\frac{\alpha_s}{6\pi}\,\big(1.479+2.855+\ldots\big)\,\big].
\end{eqnarray}
The NLO quark mass effects in the polarized decay functions can be seen to be 
large. The largest NLO quark mass correction arises in the polarized decay 
function $H^{\rm I}_{--}(b\bar b)$ where the mass correction amounts to a 
$O(200\%)$ effect compared to the leading NLO term.  

For the unpolarized decay function $H_{U+L}$ one obtains
\begin{equation}
H_{U+L}(b\bar b)=4N_cq^2\,\big[\,1.471
+\frac{\alpha_s}{6\pi}\,\big(8.874+1.219+\ldots\big)\,\big]
\end{equation}
The $O(\mu_{b})$ quark mass and radiative corrections effects increase the zero
mass Born term decay rate by $3.7\%$, i.e. one has (we take
$\alpha_s(m_Z^2)=0.115$)
\begin{equation}\label{ratecorbb}
\Gamma(\,{\rm NLO};Z \to b\bar b,O(\mu_{b}))
 =1.037\,\,\cdot\Gamma(\,{\rm Born};Z \to b\bar b,\mu_{b}=0)\,.
\end{equation}

Next we write down the angular decay distribution for
$Z(\uparrow)\to q\bar{q}$ with $Z$ polarization obtained from the production
process $e^+e^- \to Z(\uparrow)$. In terms of the $VV$, $AA$ and $VA$ structure
functions in Eq.~(\ref{vvaa}) one has
\begin{eqnarray}\label{angdistz}
W(\theta)&=&\frac38(1+\cos^2\theta)\,\,\left(
  v_f^2\,H_{U}^{VV}+a_f^2\,H_{U}^{AA}\right)\nonumber\\&&
  +\frac34\cos\theta\,\,\frac{2v_ea_e\,2v_fa_f}{v_e^2+a_e^2}\,H_{F}^{VA}
  \nonumber\\&&
  +\frac34\sin^2\theta\,\,\left(
  v_f^2\,H_{L}^{VV}+a_f^2\,H_{L}^{AA}\right).
\end{eqnarray}
Note that the electroweak parameters $v_e$ and $a_e$ do not appear in the
first and last row of Eq.~(\ref{angdistz}) because we are using normalized
density matrix elements such that $\rho_{++}+\rho_{--}=1$. The normalized
decay distribution is obtained from Eq.~(\ref{angdistz}) through 
$\widehat{W}(\theta)=W(\theta)/W$ where
$W=(v_f^2\,H_{U+L}^{VV}+a_f^2\,H_{U+L}^{AA})$.

Numerically one obtains ($m_Z=91.188\GeV$, $\alpha_s(m^2_Z)=0.115$)
\begin{equation}\label{angdistbb}
\widehat{W}^{\rm I}(\theta)\,(b\bar b)=
\frac38(1+\cos\theta)^2\Bigg\{ 
\begin{tabular}{lr}
0.570\\[-2.0ex]
0.556\\[-2.0ex]
0.559
\end{tabular}
\Bigg\}
 +\frac38(1-\cos\theta)^2\Bigg\{ 
\begin{tabular}{lr}
0.428\\[-2.0ex]
0.420\\[-2.0ex]
0.421
\end{tabular} 
\Bigg\}
+\frac34\sin^2\theta\,\Bigg\{ 
\begin{tabular}{lr}
0.002\\[-2.0ex]
0.024\\[-2.0ex]
0.020
\end{tabular} 
\Bigg\}.
\end{equation}

\begin{figure}[t]
\epsfig{figure=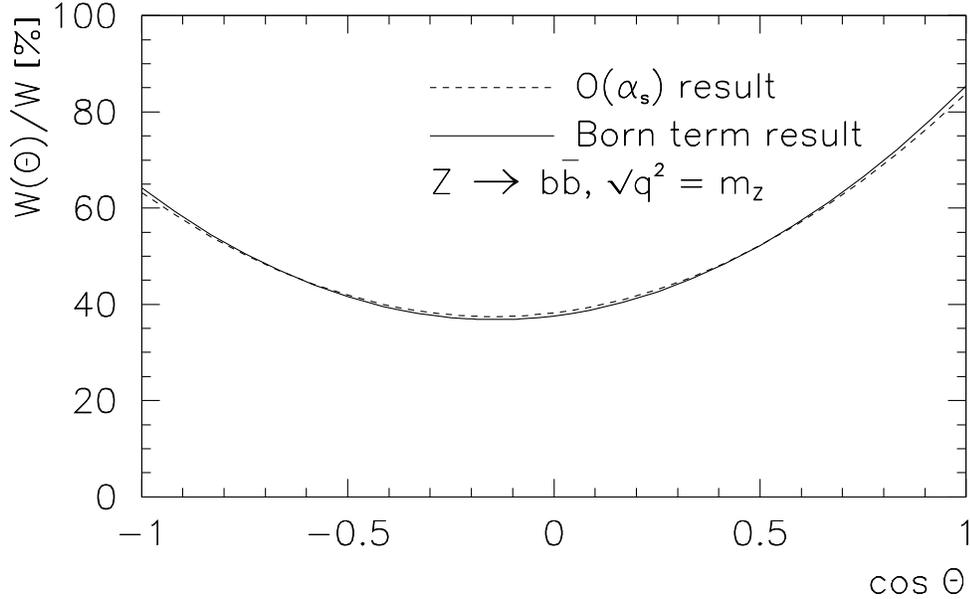, scale=0.8}
\caption{\label{wfunzb}Normalized angular decay distribution
$\widehat{W}(\theta)=W(\theta)/W$ for $Z\to b\bar b$ at LO (full line) and
NLO (dotted line), including $O(m_c^2)$ finite quark mass contributions.}
\end{figure}

Eq.~(\ref{angdistbb}) describes an upward bent parabola with unit area and a 
minimum slightly displaced to the left of $\cos\theta=0$ (see
Fig.~\ref{wfunzb}). Final state radiative corrections have a $O((1-2)\%)$
effect (on an absolute scale) on the coefficient functions of the angular
decay distribution while NLO quark mass effects contribute only at the per
mill level. In Fig.~\ref{wfunzb} we compare the $O(\mu)$ angular decay
distributions for the Born term case and the $O(\alpha_s)$ case. The NLO
corrections can be seen to make the angular decay distribution flatter.
In fact, the convexity parameter is reduced by $5.36\%$ through the radiative
corrections as Table 1 shows. The decay distribution is weighted towards the
forward hemisphere such that $A_{FB}$ is positive. It is barely visible that
the radiative corrections reduce the forward-backward asymmetry. This is born
out in Table~1 where one finds a $2.73\%$ reduction in $A_{FB}$. The minimum
is slightly shifted to the left. Quantitatively, this amounts to a $2.78\%$
effect as Table~1 shows. 

A rough description of the global effects of the radiative corrections on the
decay distribution can again be obtained by approximate formulas using the
same set of approximations as in Eq.~(\ref{rough1}-\ref{rough3}). One now 
obtains
\begin{eqnarray}
c^{\rm I}(b\bar b)&=&\frac34(1-3\rho_{00})
\left(1-\frac{\alpha_s}{6\pi}(12-2.93)\right)=0.708\,,\label{rough7}\\
A^{\rm I}_{FB}(b\bar b)&=&-\frac34(\rho_{++}-\rho_{--})\frac{2v_fa_f}{v_f^2+a_f^2}
  \left(1-\frac{\alpha_s}{6\pi}(6-1.37)\right)=0.103\,,\label{rough8}\\
\cos\theta\,\Big|_{\,\rm min}^{\rm I}(b\bar b)
  &=& \frac{(\rho_{++}-\rho_{--})}{1-3\rho_{00}}\ \frac{2v_fa_f}{v_f^2+a_f^2}
  \left(1+\frac{\alpha_s}\pi(6-1.56)\right)=-0.145\,.\label{rough9}
\end{eqnarray}
The numerical values obtained from the approximate formulas are quite close to
the relevant numbers in Table 1 (column 4; tier 3) indicating that the
approximation is quite good. The NLO finite mass effects in
Eqs.~(\ref{rough7}--\ref{rough9}) can be seen to reduce the respective
leading NLO term by $\sim 25\%$. This is in accordance with the numbers in
Table 1 and similar to what happens in the decay $W^+\to c\bar b$ (system II).

The case $Z\to c\bar c$ has to be treated separately since, apart from the
quark mass effects, one now has to use the electroweak coupling coefficients 
appropriate for up-type quarks. The numerical results are 
\begin{eqnarray}
H_U^{VV}\,(c\bar c)&=&4N_cq^2\,\big[\,1\phantom{.000}
  +\frac{\alpha_s}{6\pi}\,\big(2+0.367+\ldots\big)\,\big],\nonumber\\
H_U^{AA}\,(c\bar c)&=&4N_cq^2\,\big[\,0.999
  +\frac{\alpha_s}{6\pi}\,\big(2+0.460+\ldots\big)\,\big],\nonumber\\
H_F^{VA}\,(c\bar c)&=&4N_cq^2\,\big[\,0.999
  +\frac{\alpha_s}{6\pi}\,\big(0+0.599+\ldots\big)\,\big],\nonumber\\
H_L^{VV}\,(c\bar c)&=&4N_cq^2\,\big[\,0.001
  +\frac{\alpha_s}{6\pi}\,\big(4-0.344+\ldots\big)\,\big],\nonumber\\
H_L^{AA}\,(c\bar c)&=&4N_cq^2\,\big[\,0\phantom{.000}
  +\frac{\alpha_s}{6\pi}\,\big(4-0.307+\ldots\big)\,\big],
\end{eqnarray}
or, using Eq.~(\ref{helz})
\begin{eqnarray}
H^{\rm I}_{++}(c\bar c)&=&4N_cq^2\,\big[\,0.189
+\frac{\alpha_s}{6\pi}\,\big(1.148 -0.027+ \ldots\big)\,\big],\nonumber\\
H^{\rm I}_{00}(c\bar c)&=&4N_cq^2\,\big[\,0.000
+\frac{\alpha_s}{6\pi}\,\big(4.590-0.358 + \ldots\big)\,\big],\nonumber\\
H^{\rm I}_{--}(c\bar c)&=&4N_cq^2\,\big[\,0.957
+\frac{\alpha_s}{6\pi}\,\big(1.148+0.488 + \ldots\big)\,\big].
\end{eqnarray}
For the unpolarized decay function $H_{U+L}$ one obtains
\begin{equation}\label{ratecorcc}
H_{U+L}(c\bar c)=4N_cq^2\,\big[\,1.146+\frac{\alpha_s}{6\pi}\,\big(6.886+0.103
  +\ldots\big)\,\big]
\end{equation}
As expected, the mass corrections in the $(c \bar c)$ case are  
smaller than those in the $(b \bar b)$ case. According to~(\ref{ratecorcc}) 
the $O(\mu_c)$ quark mass and the radiative corrections effects increase the
zero mass Born term decay rate by $3.7\%$, i.e.\ one has
\begin{equation}
\Gamma(\,{\rm NLO};Z \to c\bar c,O(\mu_c))
  =1.037\,\,\cdot\Gamma(\,{\rm Born};Z \to c\bar c,\mu_c=0)\,.
\end{equation}

For the normalized angular decay distribution one finds
\begin{equation}\label{angdistcc}
\widehat{W}^{\rm I}(\theta)\,(c\bar c)
=\frac38(1+\cos\theta)^2\Bigg\{ 
\begin{tabular}{lr}
0.551\\[-2.0ex]
0.537\\[-2.0ex]
0.538
\end{tabular}
\Bigg\}
 +\frac38(1-\cos\theta)^2\Bigg\{ 
\begin{tabular}{lr}
0.449\\[-2.0ex]
0.439\\[-2.0ex]
0.440
\end{tabular} 
\Bigg\}
+\frac34\sin^2\theta\,\Bigg\{ 
\begin{tabular}{lr}
0.000\\[-2.0ex]
0.024\\[-2.0ex]
0.022
\end{tabular} 
\Bigg\}.
\end{equation}
The coefficients of the distribution~(\ref{angdistcc}) are quite similar
to those of the $(b\bar b)$ case Eq.~(\ref{angdistbb}) only that 
the NLO finite mass effects are smaller than those in the $(b\bar b)$
case. Apart from the small finite mass effects the shape of the angular decay
distribution is also affected by the difference of the factor
$2v_fa_f/(v_f^2+a_f^2)$ in the $(b \bar b)$ and $(c \bar c)$ cases which are 
given by $0.94$ and $0.67$, respectively. This is relevant for the values of
$A_{FB}$ and $c_{f}$ (see Eqs.~(\ref{rough8}) and~(\ref{rough9})). As the
relevant numbers in Table 1 show the $Z\to c\bar c$ forward-backward asymmetry
is reduced by $28.9\%$ relative to the $(b \bar b)$ case. Also the minimum of
the $c\bar c$ distribution is moved to the right by $28.4\%$ going from the
$(b\bar b)$ to the $(c\bar c)$ case (see Table~1).

%%%%%%%%%%%%%%%%%%%%%%%%%%%%%%%%%%%%%%%%%%%%%%%%%%%%%%%%%%%%%%%%%%%%%%%%%%%%%%
\section{Summary and Conclusions}
%%%%%%%%%%%%%%%%%%%%%%%%%%%%%%%%%%%%%%%%%%%%%%%%%%%%%%%%%%%%%%%%%%%%%%%%%%%%%%
We have presented $O(\alpha_s)$ results for the polarized decay functions that
describe the decay of polarized $(W,Z)$ bosons into massive quark--antiquark
pairs. NLO quark mass corrections to the polarized decay functions have
been found to be quite large. They can be as large as $200 \%$ of the leading
NLO mass term. However, these large NLO quark mass effects do not feed through
to the normalized angular decay distributions. The large NLO quark mass
effects disappear when one is dividing out the total decay function in the
normalized angular decay distribution. 

We have combined these results with information on the spin density elements
of the $(W,Z)$ bosons in the two sample production processes $t\to b+W^+$ and
$e^+e^-\to Z$ to write down explicit analytical and numerical forms of the
polar angle decay distributions. The $O(\alpha_s)$ corrections to the
polarized decay functions result in $O((1-3)\%)$ absolute changes in the
angular coefficients of the polar angle decay distributions where the bulk of
the NLO corrections come from the leading mass term. Quark mass corrections
are small but, depending on the required accuracy, are non-negligible. 

The radiative corrections make the angular decay distributions flatter for
both $W$ and $Z$ decays where the main effect comes from the  leading mass
term in the final state NLO contribution. As a measure of the flatness of the
decay distribution we have used the convexity parameter given by the second
derivative of the decay distribution. For zero mass quarks the final state
radiative corrections reduce the convexity parameters of the angular decay
distributions by $\sim(6-8)\%$. Depending on the particular case under study,
NLO quark mass effects reduce this value to $\sim(5-7)\%$. The
forward-backward asymmetry is reduced by $\sim 1.5\%$ and $\sim 3.5\%$ for $W$
and $Z$ decays, respectively, where NLO finite mass effects have reduced these
shifts by a small amount. The radiative corrections shift the position of the
maximum or minimum of the decay distributions away from zero by $\sim 7\%$ and
$\sim 3\%$ for $W$ and $Z$ decays, respectively, where again NLO finite mass
effects reduce these shifts by small amounts. The total width of the decay
$W^+\to c\bar b$ is increased by $3.8\%$ by NLO effects where NLO nonzero
quark mass effects account for $7.8\%$ of the increase. The total widths of
$Z\to b\bar b$ and $Z\to c\bar c$ are increased by $3.7\%$, where NLO mass
effects account for $13\%$ and $1.6\%$ of the increase, respectively.
 
In this paper we have only discussed polar decay correlations which probe the
diagonal spin density matrix elements of the production processes. If one
wants to probe in addition the nondiagonal spin density matrix of the gauge
bosons, one needs to involve in addition azimuthal correlations between the
momenta of the particles involved in the production/decay process. Upon
azimuthal averaging one would recover the results of the present paper
(see e.g.~\cite{Fischer:2001gp}). A comprehensive discussion of the azimuthal
correlations would form the subject of a separate publication.
 
\subsection*{Acknowledgements}
Two of us (S.G.\ and P.T.) acknowledge the support by the Estonian target
financed project No.~0180056s09 and by the Estonian Science Foundation under
grant No.~8769. S.G.\ achnowledges the support by the Deutsche
Forschungsgemeinschaft (DFG) under Grant No.~436~EST~17/1/06 and by the
Forschungszentrum of the Johannes Gutenberg--Universit\"at Mainz
``Elementarkr\"afte und Mathematische Grundlagen (EMG)''.

%%%%%%%%%%%%%%%%%%%%%%%%%%%%%%%%%%%%%%%%%%%%%%%%%%%%%%%%%%%%%%%%%%%%%%%%%%%%%%


\begin{thebibliography}{99}

\bibitem{Aaltonen:2012rz}
  T.~Aaltonen {\it et al.}  [CDF and D0 Collaborations],
  ``Combination of CDF and D0 measurements of the $W$ boson helicity in top 
   quark decays'',
  arXiv:1202.5272 [hep-ex]

\bibitem{Peters:2011an}
  Y.~Peters [D0 and CDF and Atlas and CMS Collaborations],
  ``Top Quark Properties'',
  arXiv:1112.0451 [hep-ex]

\bibitem{gkt12}
 S.~Groote, J.G.~K\"orner and P.~Tuvike, to be published

\bibitem{Brandenburg:2002xr}
  A.~Brandenburg, Z.G.~Si and P.~Uwer,
  %``QCD-corrected spin analysing power of jets in decays of polarized top
  %quarks'',
  Phys.~Lett.\ {\bf B539} (2002) 235
  %[arXiv:hep-ph/0205023].

\bibitem{Barger:1987nn}
  V.D.~Barger and R.J.N.~Phillips,
  ``Collider Physics'', in ``Frontiers in Physics'', Vol.~71, Addison Wesley,
  Redwood City, USA (1987) 592p

\bibitem{Nakamura:2010zzi}
  K.~Nakamura {\it et al.} [Particle Data Group Collaboration],
  %``Review of particle physics'',
  J.\ Phys.\ {\bf G37} (2010) 075021

\bibitem{Kane:1991bg}
  G.L.~Kane, G.A.~Ladinsky, C.P.~Yuan,
  %``Using the top quark for testing standard model polarization
  %and CP predictions'',
  Phys.\ Rev.\ {\bf D45} (1992) 124

\bibitem{Fischer:1998gsa}
  M.~Fischer, S.~Groote, J.G.~K\"orner, M.C.~Mauser and B.~Lampe,\\
  %``Polarized top decay into polarized W: t(pol.) --> W(pol.) + b at
  %O(alpha(s))'',
  Phys.~Lett.\ {\bf B451} (1999) 406
  %[arXiv:hep-ph/9811482].
  %%CITATION = PHLTA,B451,406;%%

\bibitem{Fischer:2000kx}
  M.~Fischer, S.~Groote, J.G.~K\"orner and M.C.~Mauser,
  %``Longitudinal, transverse plus and transverse minus $W$ bosons in
  %unpolarized top quark decays at O(alpha($s$) )'',
  Phys.~Rev.\ {\bf D63} (2001) 031501
  %[arXiv:hep-ph/0011075].
  %%CITATION = PHRVA,D63,031501;%%

\bibitem{Fischer:2001gp}
  M.~Fischer, S.~Groote, J.G.~K\"orner and M.C.~Mauser,
  %``Complete angular analysis of polarized top decay at O(alpha($s$) )'',
  Phys.~Rev.\ {\bf D65} (2002) 054036
  %[arXiv:hep-ph/0101322].
  %%CITATION = PHRVA,D65,054036;%%

\bibitem{Do:2002ky}
  H.S.~Do, S.~Groote, J.G.~K\"orner and M.C.~Mauser,
  %``Electroweak and finite width corrections to top quark decays into
  %transverse and longitudinal $W$ bosons'',
  Phys.~Rev.\ {\bf D67} (2003) 091501
  %[arXiv:hep-ph/0209185].
  %%CITATION = PHRVA,D67,091501;%%

\bibitem{Czarnecki:2010gb}
  A.~Czarnecki, J.G.~K\"orner and J.H.~Piclum,
  %``Helicity fractions of W bosons from top quark decays at NNLO in QCD'',
  Phys.\ Rev.\ {\bf D81} (2010) 111503
  %[arXiv:1005.2625 [hep-ph]].
  %%CITATION = PHRVA,D81,111503;%%

\bibitem{Groote:2010zf}
  S.~Groote, J.G.~K\"orner, B.~Meli\'c and S.~Prelovsek,
  %``A survey of top quark polarization at a polarized linear $e^+ e^-$ 
  %collider'',
  Phys.\ Rev.\ {\bf D83} (2011) 054018

\bibitem{Jersak:1979uv}
  J.~Jersak, E.~Laermann and P.M.~Zerwas,
  %``QCD Corrected Forward Backward Asymmetry of Quark Jets in
  %e+ e- Annihilation,''
  Phys.~Lett.\ {\bf B98} (1981) 363

\bibitem{Korner:1985dt}
  J.G.~K\"orner, G.~Schuler, G.~Kramer and B.~Lampe,
  %``Calculation of the $O(\alpha_s^2)$ Parity Violating Structure Functions 
  %in $e^+e^- \to q \bar{q}G$'',
  Z.\ Phys.\ {\bf C32} (1986) 181

\end{thebibliography}
\end{document}